\documentclass[a4paper,12pt]{article}
\usepackage{epsfig,psfig,epsf}

\usepackage{citesort}
 \normalsize
\def\du#1#2{{\left(\delta^u_{#1}\right)_{#2}}}
\def\dd#1#2{{\left(\delta^d_{#1}\right)_{#2}}}

\newcommand{\bea}{\begin{eqnarray}}
\newcommand{\eea}{\end{eqnarray}}

\begin{document}

\def\lsim{\raise0.3ex\hbox{$\;<$\kern-0.75em\raise-1.1ex\hbox{$\sim\;$}}}

\def\gsim{\raise0.3ex\hbox{$\;>$\kern-0.75em\raise-1.1ex\hbox{$\sim\;$}}}

\def\Frac#1#2{\frac{\displaystyle{#1}}{\displaystyle{#2}}}
\def\no{\nonumber\\}


\def\dofig#1#2{\centerline{\epsfxsize=#1\epsfig{file=#2, width=10cm,
height=8cm, angle=0}}}

\def\dobox#1#2{\centerline{\epsfxsize=#1\epsfig{file=#2, width=10cm,
height=5.5cm, angle=0}}}


%
\def\dofigs#1#2#3{\centerline{\epsfxsize=#1\epsfig{file=#2, width=6cm,
height=7.5cm, angle=-90}\hspace{0cm}
\hfil\epsfxsize=#1\epsfig{file=#3,  width=6cm, height=7.5cm,
angle=-90}}}
\def\dofourfigs#1#2#3#4#5{\centerline{
\epsfxsize=#1\epsfig{file=#2, width=6cm,height=7.5cm, angle=-90}
\hspace{0cm}
\hfil
\epsfxsize=#1\epsfig{file=#3,  width=6cm, height=7.5cm, angle=-90}}
 
\vspace{0.5cm}
\centerline{
\epsfxsize=#1\epsfig{file=#4, width=6cm,height=7.5cm, angle=-90}
\hspace{0cm}
\hfil
\epsfxsize=#1\epsfig{file=#5,  width=6cm, height=7.5cm, angle=-90}}
}
 
\def\dosixfigs#1#2#3#4#5#6#7{\centerline{
\epsfxsize=#1\epsfig{file=#2, width=6cm,height=7cm, angle=-90}
\hspace{-1cm}
\hfil
\epsfxsize=#1\epsfig{file=#3,  width=6cm, height=7cm, angle=-90}}
 
\centerline{
\epsfxsize=#1\epsfig{file=#4, width=6cm,height=7cm, angle=-90}
\hspace{-1cm}
\hfil
\epsfxsize=#1\epsfig{file=#5,  width=6cm, height=7cm, angle=-90}}
 
\centerline{
\epsfxsize=#1\epsfig{file=#6, width=6cm,height=7cm, angle=-90}
\hspace{-1cm}
\hfil
\epsfxsize=#1\epsfig{file=#7,  width=6cm, height=7cm, angle=-90}}
}

\def\no{\nonumber\\}
\def\slash#1{\ooalign{\hfil/\hfil\crcr$#1$}}
\def\ep{\eta^{\prime}}
\def\susy{\mbox{\tiny SUSY}}
\def\sm{\mbox{\tiny SM}}
\def\bsg{$b\to s \gamma~$}
\def\bbar{$B-\bar{B}~$}
%
\begin{titlepage}
\vspace*{-2cm}
\begin{flushright}
HIP-2005-11/TH\\
\end{flushright}
\vspace{0.1cm}
 
{\Large
\begin{center}
{\bf Supersymmetric Models and CP violation in B decays}
\end{center}
}
\vspace{.5cm}
 
\begin{center}
{E. Gabrielli$^{a,c}$, K. Huitu$^{a,b}$, and S. Khalil$^{d,e}$}
\\[5mm]
{$^{a}$\textit{Helsinki Institute of Physics,
P.O.B. 64, 00014 University of  Helsinki, Finland }}\\[0pt]
{$^{b}$\textit{Div. of HEP, Dept. of Phys.,
P.O.B. 64, 00014 University of  Helsinki, Finland}} \\[0pt]
{$^{c}$\textit{CERN PH-TH, Geneva 23, Switzerland}}\\[0pt]
{$^{d}$\textit{IPPP, University of Durham, South Rd., Durham
DH1 3LE, U.K.}}\\[0.pt]
{$^{e}$\textit{Dept. of Math., German University in Cairo - GUC,
New Cairo, El Tagamoa Al Khames, Egypt.}}\\
[0pt]
 
\begin{abstract}
In this talk CP violation in the supersymmetric models, and especially
in $B$-decays is discussed.
We review our analysis of the supersymmetric contributions to the mixing
CP asymmetries of $B\to \phi K_S$ and $B\to
\eta^{\prime} K_S$ processes. 
Both  gluino and chargino exchanges are considered in a model 
independent way by using the mass insertion approximation method.
The QCD factorization method is used, and parametrization of this 
method in terms of Wilson coefficients is presented in both 
decay modes.
Correlations between the CP asymmetries of these processes and the 
direct CP asymmetry in $b\to s \gamma$ decay are shown.
\end{abstract}

\end{center}

\end{titlepage}
 
\section{Introduction}
In the Standard Model (SM), the CP violation is due to the misalignment
of the mass and charged current interaction eigenstates.
This misalignment is represented in the CKM matrix $V_{CKM}$ \cite{KM}, 
present in the charged current interaction Lagrangian,
\bea
L^{CC}_{int}=-\frac{g_2}{\sqrt{2}}\left(\begin{array}{ccc}
\bar u_L & \bar c_L & \bar t_L \end{array}\right) \gamma_\mu
V_{CKM}\left( \begin{array}{c} d_L\\s_L\\b_L \end{array}\right)
W_\mu^+ +h.c. .
\eea
In the Wolfenstein parametrization $V_{CKM}$ is given by
\bea
V_{CKM}=\left(\begin{array}{ccc} V_{ud} & V_{us} & V_{ub} \\
 V_{cd} & V_{cs} & V_{cb} \\
 V_{td} & V_{ts} & V_{tb} \\
\end{array} \right)=
\left(\begin{array}{ccc}
1-\frac 12 \lambda^2 & \lambda & A\lambda^3 (\rho -i\eta )\\
-\lambda & 1-\frac 12 \lambda^2 & A\lambda^2 \\
A\lambda^3 (1-\rho - i\eta ) & -A\lambda^2 & 1 \end{array} \right)
+{\cal {O}} (\lambda^4) ,
\eea
where the Cabibbo mixing angle $\lambda=0.22$.
The CKM matrix is unitary, $V_{CKM}^\dagger
V_{CKM}=1=V_{CKM}V_{CKM}^\dagger$.
The unitarity conditions provide strong constraints for CP violation
in the Standard Model.
It is, however, well known that the amount of CP violation in the
Standard Model is not enough to account for the generation of the
matter-antimatter asymmetry in the universe.
Thus new sources for CP violation are expected from beyond the 
Standard Model scenarios.
{\it E.g.} in supersymmetric models a large number of new phases
emerge.
These phases are strongly constrained by electric dipole moments.

The unitarity constraints can be represented as unitarity triangles,
for which the length of the sides correspond to the products of
elements in the CKM matrix due to 
the unitarity conditions.
In the Standard Model, the angle $\beta$ in the unitary triangle
\cite{CKMfitter}, can be measured from $B$ meson decays.
The golden mode $B^0\to J/\psi K_S$ is dominated by tree contribution
and measurement of the CP asymmetries very accurately gives the
$\beta$ angle.
 
The dominant part of the decay amplitudes for $B^0\to \phi K_S,\;\eta'
K_S$
is assumed to come from the gluonic penguin,
but some contribution from the
tree level $b\to u\bar{u} s$ decay is expected.
The $|\phi\rangle$ is almost pure $|s\bar{s}\rangle$ and consequently
this decay mode corresponds also accurately, up to terms of
orders ${\cal{O}}(\lambda^2 )$,
to $\sin 2\beta$ in the SM \cite{Grossman}.
The $b\to u\bar{u} s$ tree level contribution to $B_d\to \eta' K$ was
estimated in \cite{London2}.
It was found that the tree level amplitude is less than 2\% of the
gluonic penguin amplitude.
Thus also in this mode one measures the angle $\beta$ with
a good precision in the SM.
Therefore, it is expected that NP contributions to the CP asymmetries in
$B^0\to \phi K_S,\;\eta' K_S$ decays are more significant than in
$B^0\to J/\psi K_S$ and can compete with the SM one.

$B$ physics is a natural framework to test the beyond the Standard Model CP
violation effects.  
It is clear that ultimately one needs to test the Standard Model
with three generations and that large statistics is needed to achieve 
the small effects of CP violation. 
Flow of interesting data has been provided in recent years by 
the B-factories.

In this talk we will not assume any particular model for CP violation,
but consider a general SUSY model.
The talk is based on papers \cite{GHK,CGHK}, where a comparative study
of SUSY contributions from
chargino and gluino to $B\rightarrow \phi K$ and
$B\rightarrow \eta^{\prime} K$ processes in naive factorization (NF) 
and QCD factorization (QCDF) approaches is done.
We also analyzed in \cite{GHK} the branching ratios of these decays and
investigated their correlations with CP asymmetries.
The correlations between CP asymmetries of these
processes and the direct CP asymmetry in $b\to s \gamma$ decay
\cite{ACPbsg}
is also discussed in \cite{GHK}.

In the analysis the mass insertion method (MIA) \cite{mia} is used.
Denoting by $\Delta^q_{AB}$ the off--diagonal terms
in the sfermion $(\tilde{q}=\tilde{u},\tilde{d})$
mass matrices for the up and down, respectively,
where $A,B$ indicate chirality couplings to
fermions $A,B=(L,R)$, the A--B squark propagator can be expanded as
\begin{equation}
\langle \tilde{q}^a_A \tilde{q}^{b*}_B \rangle =
i \left(k^2{\bf 1} - \tilde{m}^2 {\bf 1}
- \Delta_{AB}^q\right)^{-1}_{ab} \simeq \frac{i \delta_{ab}}
{k^2 - \tilde{m}^2} +
~\frac{i (\Delta_{AB}^q)_{ab}}{(k^2 -\tilde{m}^2)^2} +
{\cal O}(\Delta^2) ,
\end{equation}
where $q=u,d$ selects up or down sector, respectively,
$a,b=(1,2,3)$ are flavor indices, ${\bf 1}$
is the unit matrix, and $\tilde{m}$ is the average squark mass.
As we will see in the following, it is convenient to parametrize
this expansion in terms of the dimensionless quantity
$(\delta_{AB}^q)_{ab} \equiv (\Delta^q_{AB})_{ab}/\tilde{m}^2$.

New physics (NP) could in principle affect the $B$ meson decay
by means of a new source of CP violating phase in the corresponding
amplitude. In general this phase is different from the corresponding SM
one.
If so, then deviations on CP asymmetries from SM
expectations can be sizeable, depending on the relative magnitude of SM
and NP amplitudes. For instance, in the SM the
$B\to \phi K_S$ decay amplitude
is generated at one loop and therefore it is
very sensitive to NP contributions.
In this respect, SUSY models with non minimal flavor structure and new
CP violating phases in the squark mass matrices,
can easily generate large deviations in the
$B\to \phi K_S$  asymmetry.
 
The time dependent CP asymmetry for $B\to \phi K_S$ can
be described by
\begin{eqnarray}
a_{\phi K_S}(t)&=&\frac{\Gamma (\overline{B}^0(t)\to\phi K_S)-\Gamma
(B(t)\to\phi K_S)} {\Gamma (\overline{B}^0(t)\to\phi K_S)+\Gamma (B(t)
\to\phi K_S)}
= C_{\phi K_S}\cos\Delta M_{B_d}t+S_{\phi K_S}\sin\Delta
M_{B_d}t,\nonumber\\
\label{asym_phi}
\end{eqnarray}
where $C_{\phi K_S}$ and $S_{\phi K_S}$ represent
the direct  and the mixing CP asymmetry, respectively and they are given
by
\begin{equation}
C_{\phi K_S}=
\frac{|\overline{\rho}(\phi K_S)|^2-1}{|\overline{\rho}(\phi K_S)|^2+1},
\ \
S_{\phi K_S}=\frac{2Im \left[\frac{q}{p}~\overline{\rho}(\phi
K_S)\right]}
{|\overline{\rho}(\phi K_S)|^2+1}.
\label{S_PHI}
\end{equation}
The parameter $\overline{\rho}(\phi K_S)$ is defined by
\begin{equation}
\overline{\rho} (\phi K_S)=\frac{\overline{A}(\phi K_S)}{A(\phi K_S)}.
\end{equation}
where $\overline{A}(\phi K_S)$ and $A(\phi K_S)$ are
the decay amplitudes of $\overline{B}^0$ and $B^0$  mesons,
respectively.
Here, the mixing parameters $p$ and $q$ are defined by
$|B_1\rangle =p|B\rangle +q|\overline{B}^0\rangle ,
\ \ |B_2\rangle =p|B\rangle -q|\overline{B}^0\rangle$
where $|B_{1(2)}\rangle$ are mass eigenstates of $B$ meson.
The ratio of the mixing parameters is given by
\begin{equation}
\frac{q}{p}=-e^{-2i\theta_d}\frac{V_{tb}^*V_{td}}{V_{tb}V_{td}^*},
\end{equation}
where $\theta_d$ represent any SUSY contribution
to the $B-\bar{B}^0$ mixing angle.
Finally, the above amplitudes can be written in terms of the matrix
element
of the $\Delta B=1$ transition as
\begin{equation}
\overline{A}(\phi K_S)=
\langle \phi K_S| H_{eff}^{\Delta B=1}|
\overline{B}^0\rangle,
\ \ \
A(\phi K_S)=
\langle \phi K_S| \left(H_{eff}^{\Delta B=1}\right)^{\dag}|B^0\rangle.
\end{equation}

Results by BaBar and Belle collaboration 
have been announced in \cite{giorgi,sakai}.
The experimental value of the indirect CP asymmetry parameter
for $B^0\to J/\psi K_S$ is given by \cite{giorgi,sakai}
\bea
S_{J/\psi K_S} =0.726\pm 0.037,
\label{Spsi}
\eea
which agrees quite well with the SM prediction $0.715_{-0.045}^{+0.055}$
\cite{AB}.
Results on the corresponding $\sin{2\beta}$
extracted for $B^0\to \phi K_S$ process is as follows
\cite{giorgi,sakai}
\bea
S_{\phi K_S}&=&0.50\pm 0.25^{+0.07}_{-0.04}\;\;
({\rm BaBar}),\nonumber\\
&=&0.06\pm 0.33 \pm 0.09\;\; ({\rm Belle})\, ,
\label{Sphi}
\eea
where the first errors are statistical and the second systematic.
As we can see from Eq.(\ref{Sphi}),
the relative central values are different. BaBar results
\cite{giorgi} are more
compatible with SM predictions, while Belle measurements
\cite{sakai} show
a deviation from the $c\bar{c}$ measurements of about $2\sigma$.
Moreover, the average $S_{\phi K_S}=0.34 \pm 0.20$
displays 1.7$\sigma$ deviation from Eq.(\ref{Spsi}).
 
Furthermore, the most recent measured CP asymmetry
in the $B^0\to \eta^{\prime} K_S$ decay is found by BaBar \cite{giorgi}
and Belle \cite{sakai} collaborations as
\bea
S_{\eta^{\prime} K_S}&=&0.27\pm 0.14\pm 0.03 \;\; ({\rm BaBar})
\nonumber\\
&=&0.65\pm 0.18\pm 0.04 \;\; ({\rm Belle}),
\label{Seta}
\eea
with an average $S_{\eta^{\prime} K_S}= 0.41 \pm 0.11$,
which shows a 2.5$\sigma$  discrepancy from Eq.~(\ref{Spsi}).
 
It is interesting to note that the results on s-penguin modes
from both experiments differ from the value extracted from the
$c\bar{c}$ mode ($J/\psi$), BaBar by
2.7$\sigma$ and Belle by 2.4$\sigma$ \cite{giorgi,sakai}.
At the same time the experiments agree with each other, and even
the central values are quite close:
\bea
0.42\pm0.10\;\; {\rm BaBar},\;\;\; 0.43^{+0.12}_{-0.11}\;\;{\rm
  Belle}.
\nonumber
\eea
 
On the other hand, the experimental measurements of the branching
ratios of $B^0\to \phi K^0$ and $B^0\to \eta^{\prime} K^0$ at BaBar
\cite{BR_babar}, Belle \cite{BR_belle},
and CLEO \cite{BR_cleo} lead to the following averaged  results
\cite{hfag} :
\begin{eqnarray}
BR(B^0\to \phi K^0) &=& (8.3^{+1.2}_{-1.0}) \times 10^{-6},
\label{BRphi}\\
BR(B^0\to \eta^{\prime} K^0) &=& (65.2^{+6.0}_{-5.9} ) \times 10^{-6}.
\label{BReta}
\end{eqnarray}
From theoretical side, the SM predictions
for $BR(B\to \phi K)$
are in good agreement with Eq.(\ref{BRphi}),
while showing a large discrepancy, being experimentally two to five
times
larger, for
$BR(B\to \eta^{\prime} K)$ in Eq.(\ref{BReta})
\cite{BR_eta_SM}.
This discrepancy is not new and it
has created a growing interest in the subject.
However, since it is observed only in $B\to \eta^{\prime} K$ process,
mechanisms
based on the peculiar structure of $\eta^{\prime}$ meson, such as
intrinsic charm and gluonium content, have been investigated to solve 
the puzzle.
Correlations with branching ratios have been discussed in \cite{GHK}.

\section{SUSY contributions to $B\to \phi(\eta')K$ decay}
%
We first consider the supersymmetric effect in
the non-leptonic $\Delta B=1$ processes.
Such an effect could be a probe for any testable
SUSY implications in CP violating experiments.
The most general effective Hamiltonian $H^{\Delta B=1}_{\rm eff}$
for the non-leptonic $\Delta B=1$  processes can be expressed via 
the Operator
Product Expansion (OPE) as \cite{bbl}
\bea
H^{\Delta B=1}_{\rm eff}&=& \left\{\frac{G_F}{\sqrt{2}}
\sum_{p=u,c} \lambda_p ~\left( C_1 Q_1^p + C_2 Q_2^p +
\sum_{i=3}^{10} C_i Q_i + C_{7\gamma}
Q_{7\gamma} + C_{8g} Q_{8g} \right) \right\}
\nonumber\\
&+&\left\{Q_i\to \tilde{Q}_i\, ,\, C_i\to \tilde{C}_i\right\}
\;,
\label{Heff}
\eea
where $\lambda_p= V_{pb} V^{\star}_{p s}$,
with $V_{pb}$ the unitary CKM matrix elements satisfying the unitarity
triangle relation
$\lambda_t+\lambda_u+\lambda_c=0$, and $C_i\equiv C_i(\mu_b)$ are
the Wilson coefficients at low energy scale
$\mu_b\simeq {\cal O}(m_b)$.
The basis $Q_i\equiv Q_i(\mu_b)$
is given by the relevant local operators
renormalized at the same scale $\mu_b$, namely
\begin{eqnarray}
Q^p_2 &=& (\bar p b )_{V-A}~~ (\bar s p)_{V-A}\;,~~~~~~~~~~~~~~~
Q^p_1 = (\bar p_{\alpha} b_{\beta})_{V-A}~~ (\bar s_{\beta}
p_{\alpha})_{V-A}\;
\nonumber\\
Q_3 &=& (\bar s b )_{V-A}~~ \sum_q (\bar q q)_{V-A}\;,~~~~~~~~~~~
Q_4 = (\bar s_{\alpha} b_{\beta})_{V-A}~~ \sum_q (\bar q_{\beta}
q_{\alpha})_{V-A}\;,
\nonumber\\
Q_5 &=& (\bar s b )_{V-A}~~ \sum_q (\bar q q)_{V+A}\;,~~~~~~~~~~~
Q_6 = (\bar s_{\alpha} b_{\beta} )_{V-A}~~ \sum_q (\bar q_{\beta}
q_{\alpha})_{V+A}\;,
\nonumber\\
Q_7 &=& (\bar s b )_{V-A}~~ \sum_q\frac{3}{2}  e_q (\bar q
q)_{V+A}\;,~~~~~~
Q_8 = (\bar s_{\alpha} b_{\beta} )_{V-A}~~
\sum_q \frac{3}{2} e_q (\bar q_{\beta} q_{\alpha})_{V+A}\;,
\nonumber\\
Q_9 &=& (\bar s b )_{V-A}~~
\sum_q \frac{3}{2} e_q (\bar q q)_{V-A}\;,~~~~~~
Q_{10} = (\bar s_{\alpha} b_{\beta} )_{V-A}~~
\sum_q \frac{3}{2} e_q (\bar q_{\beta} q_{\alpha})_{V-A}\;,
\nonumber\\
Q_{7\gamma} &=& \frac{e}{8\pi^2} m_b \bar s \sigma^{\mu\nu}
(1+ \gamma_5) F_{\mu \nu} b\;,~~~~~~~
Q_{8g}= \frac{g_s}{8\pi^2} m_b \bar s_{\alpha} \sigma^{\mu\nu}
(1+ \gamma_5) G^A_{\mu \nu} t^A_{\alpha\beta} b_{\beta}\;.
\label{Qbasis}
\end{eqnarray}
Here $\alpha$ and $\beta$ stand for color indices, and
$t^A_{\alpha\beta}$ for
the $SU(3)_c$ color matrices,
$\sigma^{\mu\nu}=\frac{1}{2}i[\gamma^{\mu},\gamma^{\nu}]$. Moreover,
$e_q$ are quark electric charges in unity of $e$,
$(\bar q q)_{V \pm A}\equiv \bar q \gamma_\mu
(1 \pm \gamma_5) q$, and $q$ runs over $u$, $d$, $s$, and $b$ quark
labels.
In the SM only the first part
of right hand side of Eq.(\ref{Heff}) (inside first curly brackets)
containing operators $Q_i$ will contribute, where
$Q^p_{1,2}$ refer to the current-current operators,
$Q_{3-6}$ to the QCD penguin operators,
and $Q_{7-10}$ to the electroweak penguin operators, while $Q_{7\gamma}$
and
$ Q_{8g}$ are the magnetic and the chromo-magnetic dipole operators,
respectively.
In addition, operators $\tilde{Q}_{i}\equiv \tilde{Q}_{i}(\mu_b)$
are obtained from $Q_{i}$ by the chirality exchange
$(\bar{q}_1 q_2)_{V\pm A} \to (\bar{q}_1 q_2)_{V\mp A}$.
Notice that in the SM the coefficients $\tilde{C}_{i}$ identically
vanish due to the V-A structure of charged weak currents,
while in MSSM they can
receive contributions from both chargino and gluino exchanges
\cite{bbmr,ggms}.

As mentioned, we calculated in \cite{CGHK} the chargino contribution
to the Wilson coefficients in the MIA approximation.
In MIA framework, one chooses
a basis (called super-CKM basis) where the couplings of
fermions and sfermions to neutral
gaugino fields are flavor diagonal.
In this basis, the interacting Lagrangian involving charginos is given
by
\bea
\mathcal{L}_{q\tilde{q}\tilde{\chi}^+} &=& - g~
\sum_k \sum_{a,b}~ \Big(~ V_{k1}~
K_{ba}^*~ \bar{d}_L^a~ (\tilde{\chi}^+_k)^*~ \tilde{u}^b_L -  U_{k2}^*~
(Y_d^{\mathrm{diag}} . K^+)_{ab}~ \bar{d}_R^a~ (\tilde{\chi}^+_k)^*~
\tilde{u}^b_L \nonumber\\ && - V_{k2}~ (K.Y_u^{\mathrm{diag}})_{ab}~
\bar{d}_L^a~ (\tilde{\chi}^+_k)^*~ \tilde{u}^b_R~ \Big),
\label{vertices}
\eea
where $q_{R,L}=\frac{1}{2}(1\pm \gamma_5)q$, and
contraction of color and Dirac indices is understood.
Here $Y_{u,d}^{\mathrm{diag}}$ are the diagonal Yukawa matrices, and
$K$ stands for the CKM matrix.
The indices $a,b$ and $k$ label
flavor and chargino mass eigenstates, respectively, and
$V$, $U$ are the chargino mixing matrices defined by
\begin{equation}
U^* M_{\tilde{\chi}^+} V^{-1} = \mathrm{diag}(m_{\tilde{\chi}_1^+},
m_{\tilde{\chi}_2^+}),~ \mathrm{and}~ M_{\tilde{\chi}^+} = \left(
\begin{array}{cc}
M_2 & \sqrt{2} m_W \sin \beta \\
\sqrt{2} m_W \cos \beta & \mu
\end{array}\right) \;,
\label{chmatrix}
\end{equation}
where $M_2$ is the weak gaugino mass, $\mu$ is the supersymmetric
Higgs mixing term, and $\tan\beta$ is the ratio of the
vacuum expectation value (VEV) of the up-type Higgs to the VEV
of the down-type Higgs\footnote{This $\tan\beta$ should not be
confused with the angle $\beta$ of the unitarity triangle.} .
As one can see from Eq.(\ref{vertices}),
the higgsino couplings are suppressed by Yukawas of the light quarks,
and
therefore they are negligible, except for the stop--bottom
interaction which is directly enhanced by the top Yukawa ($Y_t$).
In our analysis we neglect the higgsino contributions
proportional to the Yukawa couplings of light quarks
with the exception of the bottom Yukawa $Y_b$,
since its effect could be enhanced by large $\tan{\beta}$.
However, it is easy to show that this vertex cannot
affect dimension six operators of the effective Hamiltonian
for $\Delta B=1$ transitions (operators $Q_{i=1-10}$ in
Eq.(\ref{Heff}))
and only interactions involving left down quarks will contribute.
On the contrary, contributions proportional to bottom Yukawa $Y_b$ enter
in the Wilson coefficients of dipole operators ($C_{7\gamma}$, $C_{8g}$)
due to the chirality flip of $b\to s \gamma$ and $b\to s g$ transitions.

%
%
%
The calculation of $B\to \phi(\eta^{\prime}) K$ decays involves the
evaluation of the hadronic
matrix elements of related operators in the effective Hamiltonian,
which is the most
uncertain part of this calculation.
In the limit in which $m_b \gg \Lambda_{QCD}$ and neglecting QCD
corrections in $\alpha_s$,
the hadronic matrix elements of B meson decays in two mesons
can be factorized, for example for $B\to M_1 M_2$, in the form
\begin{equation}
\langle M_1 M_2|Q_i|\bar{B}^0\rangle =\langle M_1|j_1|\bar{B}^0\rangle
\times \langle M_2|j_2|0\rangle
\end{equation}
where $M_{1,2}$ indicates two generic mesons,
$Q_i$ are local four fermion operators of the effective Hamiltonian in
Eq.(\ref{Heff}), and $j_{1,2}$ represent bilinear quark currents.
Then, the final results can be usually parametrized
by the product of the decay constants and the transition form factors.
This approach is known as naive factorization (NF) \cite{naive,ali}.
In QCDF the hadronic
matrix element for $B \to M K$ with $M=\phi, \eta^{\prime}$
in the heavy quark limit $m_b \gg \Lambda_{QCD}$ can be written as
\cite{BBNS}
\begin{equation}
\langle M K  \vert Q_i \vert B \rangle_{QCDF} =
\langle M K \vert Q_i \vert B \rangle_{NF}.
\left[ 1+ \sum_n r_n \alpha_S^n +
{\mathcal O}\left(\frac{\Lambda_{QCD}}{m_b} \right) \right],
\end{equation}
where $\langle M K\vert Q_i \vert B \rangle_{NF}$ denotes
the NF results. The second
and third term in the bracket represent the radiative
corrections in $\alpha_S$ and $\Lambda_{QCD}/m_b$ expansions, respectively.
Notice that, even though at higher order in $\alpha_s$
the simple factorization is broken, these corrections can be calculated
systematically in terms of short-distance coefficients and meson
light-cone
distribution functions.
 
In the QCD factorization method \cite{BBNS,BN},
the decay amplitudes of $B \to \phi(\eta^{\prime})K$
can be expressed as
\begin{equation}
\mathcal{A}\left(B\to \phi(\eta^{\prime})K\right) =
\mathcal{A}^f\left(B\to \phi(\eta^{\prime}) K\right)
+ \mathcal{A}^a\left(B\to \phi(\eta^{\prime}) K\right),
\end{equation}
where
\begin{equation}
\mathcal{A}^f\left(B\to \phi(\eta^{\prime}) K\right) =
\frac{G_F}{\sqrt{2}}\sum_{p=u,c}
\sum_{1=1}^{10} V_{pb}V_{ps}^*~ a_i^{\phi(\eta^{\prime})} 
\langle \phi(\eta^{\prime}) K 
\vert Q_i \vert B \rangle_{NF},
\label{nf-ampl}
\end{equation}
and
\begin{equation}
\mathcal{A}^a\left(B\to \phi(\eta^{\prime}) K\right) =
\frac{G_F}{\sqrt{2}}f_B f_{K} f_{\phi}
\sum_{p=u,c} \sum_{i=1}^{10} V_{pb}V_{ps}^*~ b_i^{\phi(\eta^{\prime})}.
\label{ann-ampl}
\end{equation}
The first term $\mathcal{A}^f\left(B\to \phi(\eta^{\prime}) K\right)$
includes vertex corrections, penguin corrections and hard spectator
scattering contributions which are involved in the parameters
$a_i^{\phi(\eta^{\prime})}$. The other term
$\mathcal{A}^a\left(B\to \phi(\eta^{\prime}) K\right)$
includes  the weak annihilation contributions which are absorbed in the
parameters $b_i^{\phi(\eta^{\prime})}$.
However, these contributions contain infrared divergences, and the
subtractions of these divergences are
usually parametrized as \cite{BN}
\begin{equation}
\int_0^1 \frac{d x}{x} \to X_{H,A} \equiv
\left(1 + \rho_{H,A} e^{i \phi_{H,A}} \right)
\ln\left(\frac{m_B}{\Lambda_{QCD}}\right),
\label{paramXAH}
\end{equation}
where $\rho_{H,A}$ are free parameters expected to be of order
of $\rho_{H,A} \simeq {\cal O}(1)$, and $\phi_{H,A} \in [0,2\pi]$.
As already discussed in Ref.\cite{BN}, if one does not require fine
tuning
of the annihilation phase $\phi_A$, the $\rho_A$ parameter gets an upper
bound
from measurements on branching ratios, which is of order of
$\rho_A \lsim 2$. Clearly, large values of $\rho_A$
are still possible, but in this case strong fine
tuning in the phase $\phi_A$ is required.
However, assumptions of very large values of $\rho_{H,A}$, which
implicitly means large contributions from
hard scattering and weak annihilation diagrams, seem to be quite
unrealistic.
In \cite{GHK} we assumed $\rho<2$.

\section{{\bf CP asymmetry in $B\to {\phi K_S}$ and in
$B\to {\eta' K_S}$}}
\begin{figure}[tpb]  
\begin{center}
\mbox{\epsfig{file=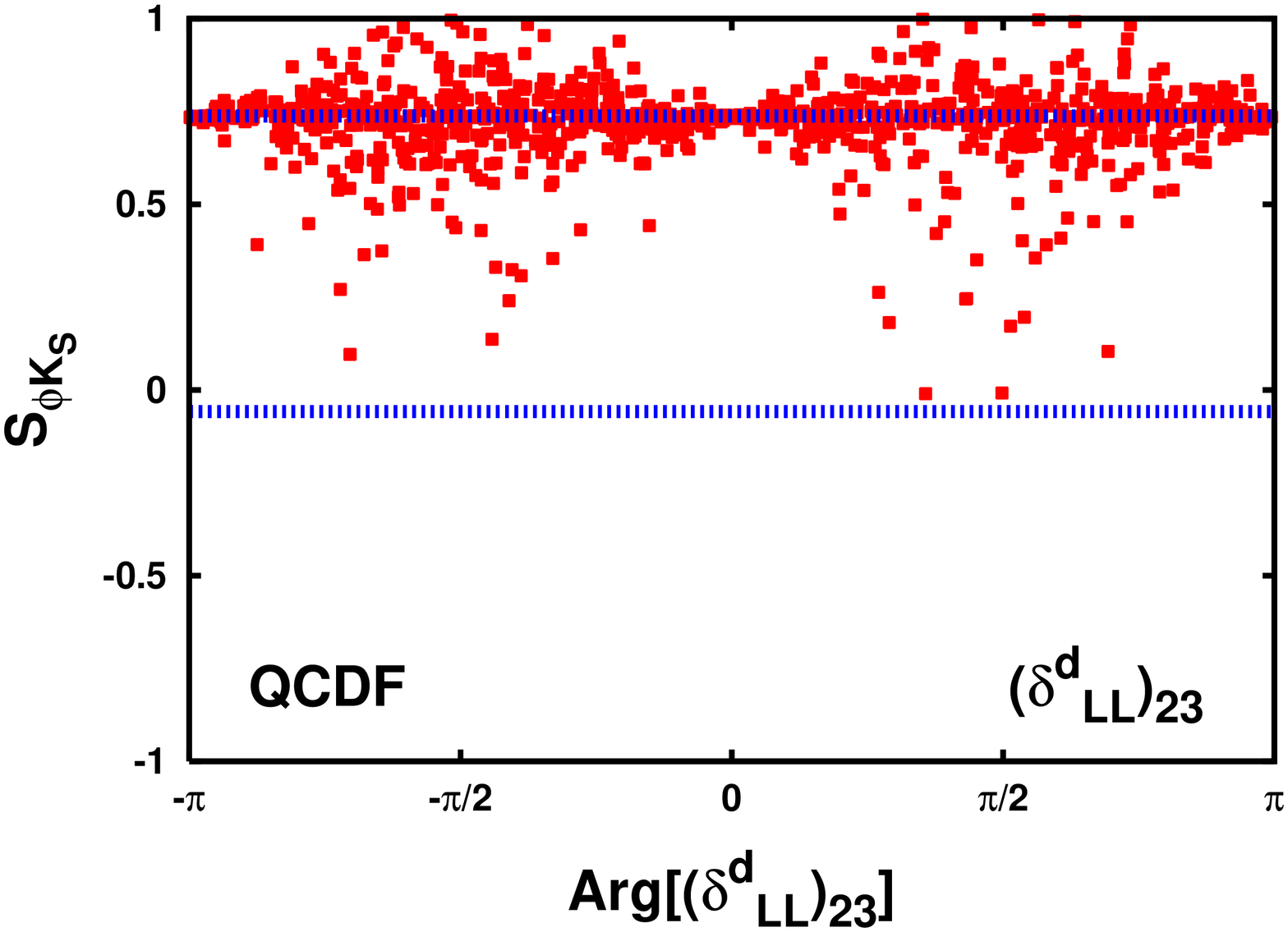, width=6truecm,
height=7truecm, angle=-90}}
\mbox{\epsfig{file=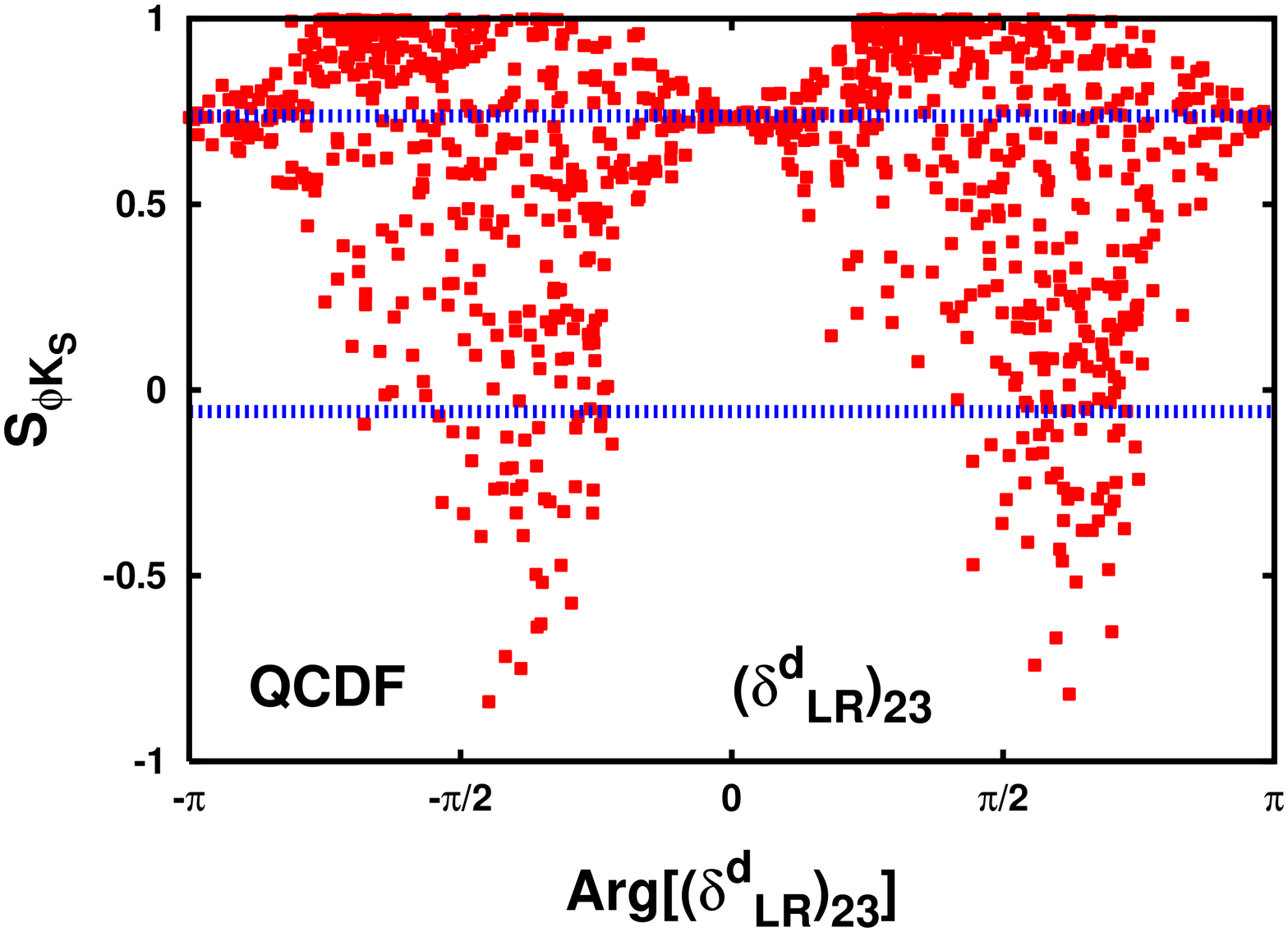, width=6truecm,
height=7truecm, angle=-90}}
\end{center}
\caption{\small $S_{\Phi K_S}$ as a function of arg[$\dd{LL}{23}$]
(left) and  arg[$\dd{LR}{23}$] (right) with  gluino
contribution of one mass insertion.
The region inside the two horizontal lines
corresponds to the allowed experimental region at $2\sigma$ level.}
\label{Fig41}
\end{figure}

In order to simplify our analysis, it is useful to parametrize the SUSY 
effects by introducing the ratio of SM and SUSY amplitudes as follows
\bea
\left(\frac{A^{\susy}}{A^{\sm}}\right)_{\phi K_S}&\equiv&
R_{\phi}~ e^{i\theta_{\phi}}~e^{i\delta_{\phi}}
\\
\label{ratioPHI}
\eea
and analogously for the $\eta^{\prime} K_S$ decay mode
\bea
\left(\frac{A^{\susy}}{A^{\sm}}\right)_{\eta^{\prime} K_S}&\equiv&
R_{\eta^{\prime}}~ e^{i\theta_{\eta^ {\prime}}}~e^{i\delta_{\eta^ {\prime}}}
\label{ratioETA}
\eea
where $R_i$ stands for the corresponding absolute values of 
$|\frac{A^{\susy}}{A^{\sm}}|$, 
the angles $\theta_{\phi,~\eta^{\prime}}$
 are the corresponding SUSY CP violating phase,
and $\delta_{\phi,~\eta^ {\prime}}=\delta^{SM}_{\phi,~\eta^ {\prime}}-
\delta^{SUSY}_{\phi,~\eta^ {\prime}}$ 
parametrize the strong (CP conserving) phases.
In this case, the mixing CP asymmetry $S_{\phi K_S}$ in Eq.(\ref{asym_phi}) 
takes the following form 
\bea
S_{\phi K_S}&=&\Frac{\sin 2 \beta +2 R_{\phi} 
\cos \delta_{\phi} \sin(\theta_{\phi}+2 \beta)+
R_{\phi}^2 \sin (2 \theta_{\phi}+2 \beta)}{1+ 2 R_{\phi} 
\cos \delta_{\phi} \cos\theta_{\phi} +R_{\phi}^2}.
\label{cpmixing_phi}
\eea
\vspace{0.5in}
and analogously for $B\to \eta^{\prime} K_S$
\bea
S_{\eta^{\prime} K_S}&=&\Frac{\sin 2 \beta +2 R_{\eta^{\prime}} 
\cos \delta_{\eta^{\prime}} \sin(\theta_{\eta^{\prime}}+2 \beta)+
R_{\eta^{\prime}}^2 \sin (2 \theta_{\eta^{\prime}}+2 \beta)}
{1+ 2 R_{\eta^{\prime}} 
\cos \delta_{\eta^{\prime}} \cos\theta_{\eta^{\prime}} +
R_{\eta^{\prime}}^2}.
\label{cpmixing_eta}
\eea

\begin{figure}[tpb]
\begin{center}
\mbox{\epsfig{file=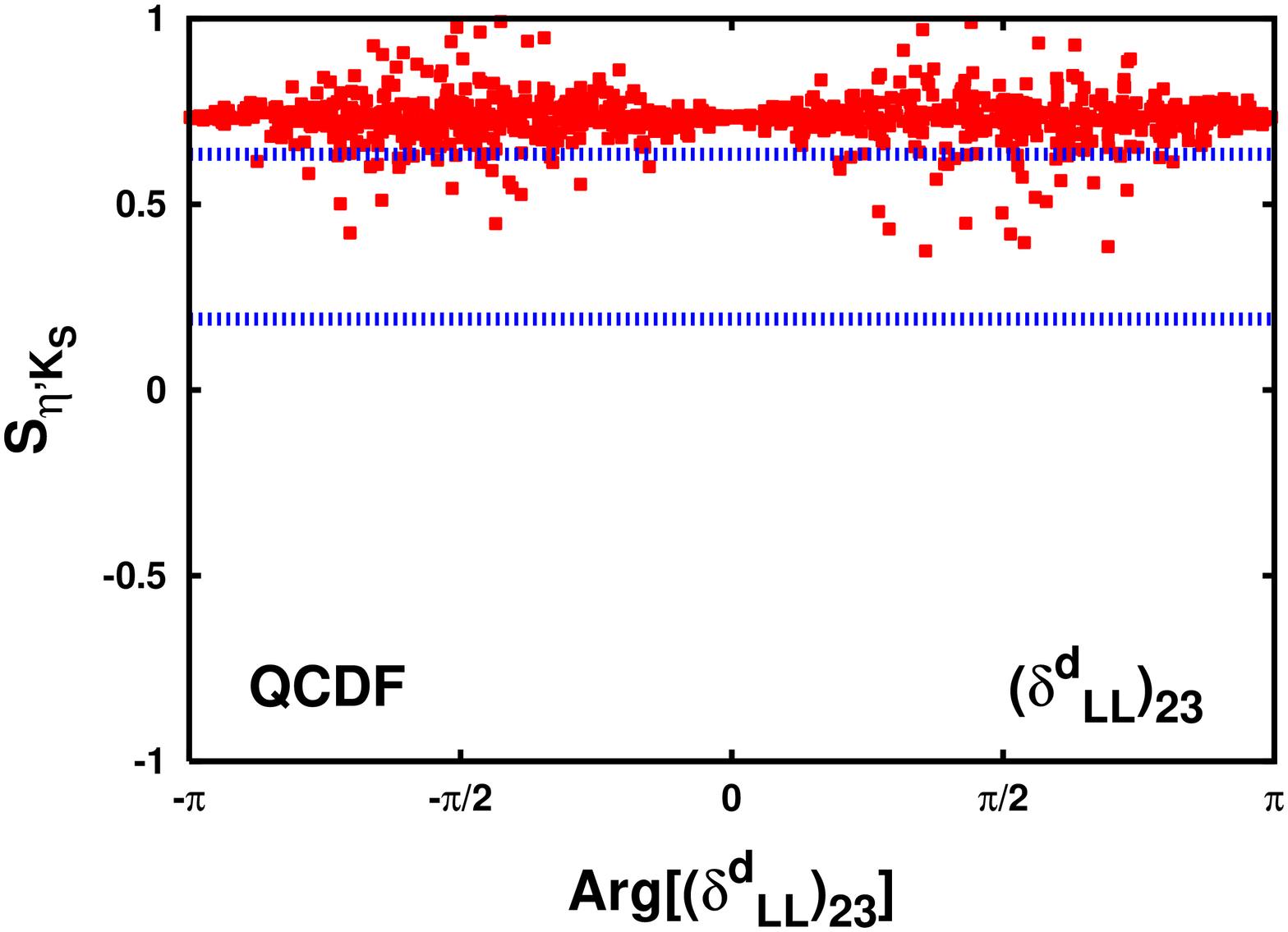, width=6truecm,
height=7truecm, angle=-90}}
\mbox{\epsfig{file=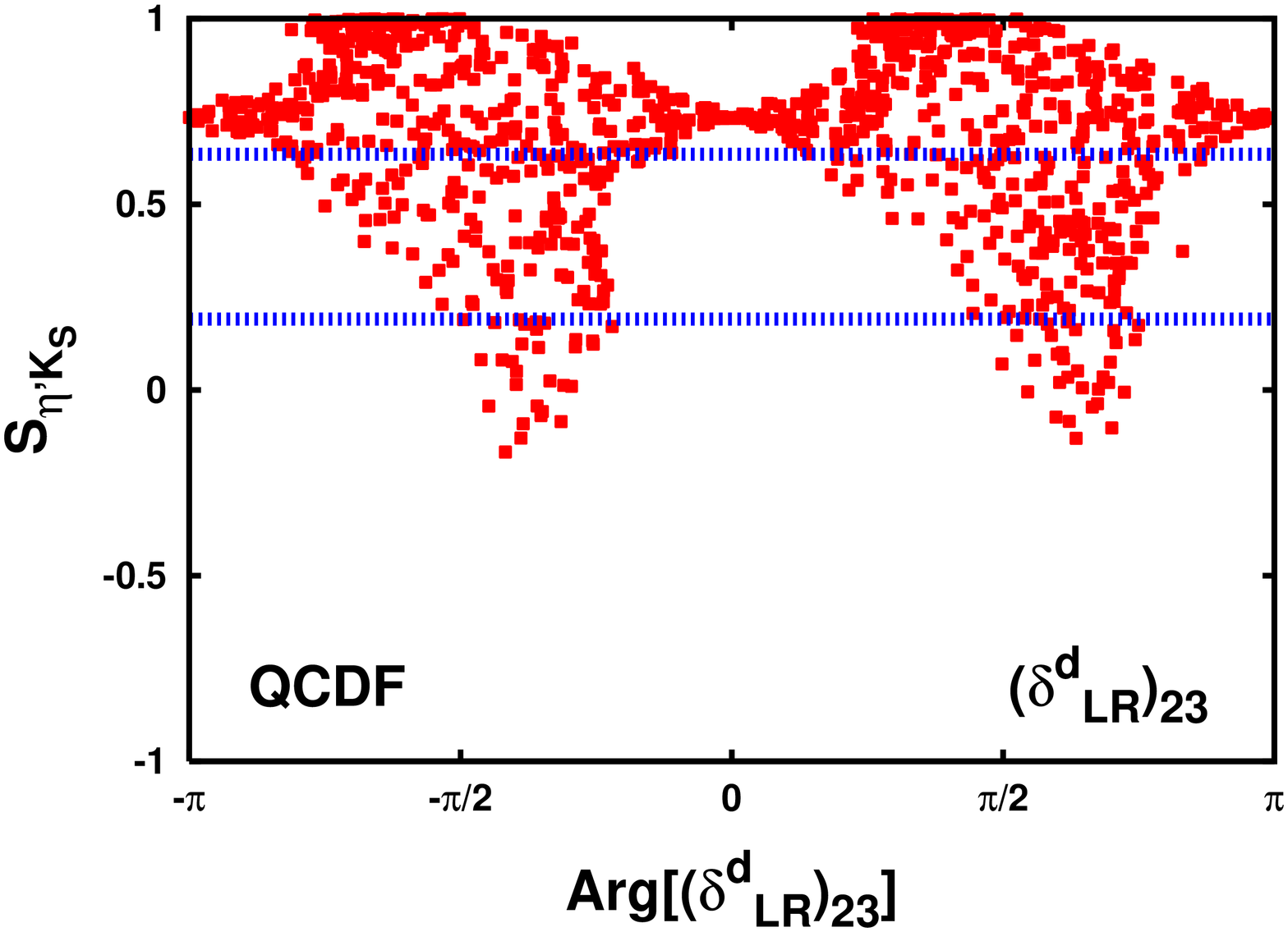, width=6truecm,
height=7truecm, angle=-90}}
\end{center}
\caption{\small As in Fig.~\ref{Fig41}, but for
$S_{\eta^{\prime} K_S}$.}
\label{Fig43}
\end{figure}

Our numerical results for the gluino contributions to CP asymmetry
 $S_{\Phi K_S}$  are presented in Fig.~\ref{Fig41}, and to CP asymmetry
 $S_{\eta' K_S}$  are presented in Fig.~\ref{Fig43}. 
In all the plots, regions inside the horizontal lines indicate
the allowed $2\sigma$ experimental range.
In the plots only one mass insertion per time is taken active,
in particular this means that we scanned over $|\dd{LL}{23}|<1$ 
and $|\dd{LR}{23}|<1$.
Then, $S_{\Phi (\eta') K_S}$ is plotted versus $\theta_{\phi}$, which in the
case of one dominant mass insertion should be identified here as
$\theta_{\phi}={\rm arg}[(\delta_{AB}^d)_{ij}]$.

We have scanned over the relevant SUSY parameter space, 
in this case the average squark mass $\tilde{m}$ and gluino
mass $m_{\tilde{g}}$, assuming SM central values \cite{PDG}.
Moreover, we require that the SUSY spectra 
satisfy the present experimental lower mass bounds \cite{PDG}. 
In particular, $m_{\tilde{g}}> 200$ GeV,
$\tilde{m} > 300$ GeV. In addition, we impose that the
branching ratio (BR) of \bsg 
and the \bbar mixing are satisfied
at  95\% C.L. \cite{bsgmeas}, namely 
$2\times 10^{-4}\le BR(b\to s\gamma) < 4.5\times 10^{-4}$.
Then, the allowed ranges for $|\dd{LL}{23}|$ 
and $|\dd{LR}{23}|$ are obtained by 
taken into account the above constraints on \bsg and \bbar mixing.
We have also scanned over the 
full range of the parameters $\rho_{A,H}$ and $\phi_{A,H}$  in 
$X_A$ and $X_H$, respectively, as defined in  Eq.(\ref{paramXAH}).

\vspace{0.1cm}
 
\begin{figure}[tpb]
\begin{center}
\mbox{\epsfig{file=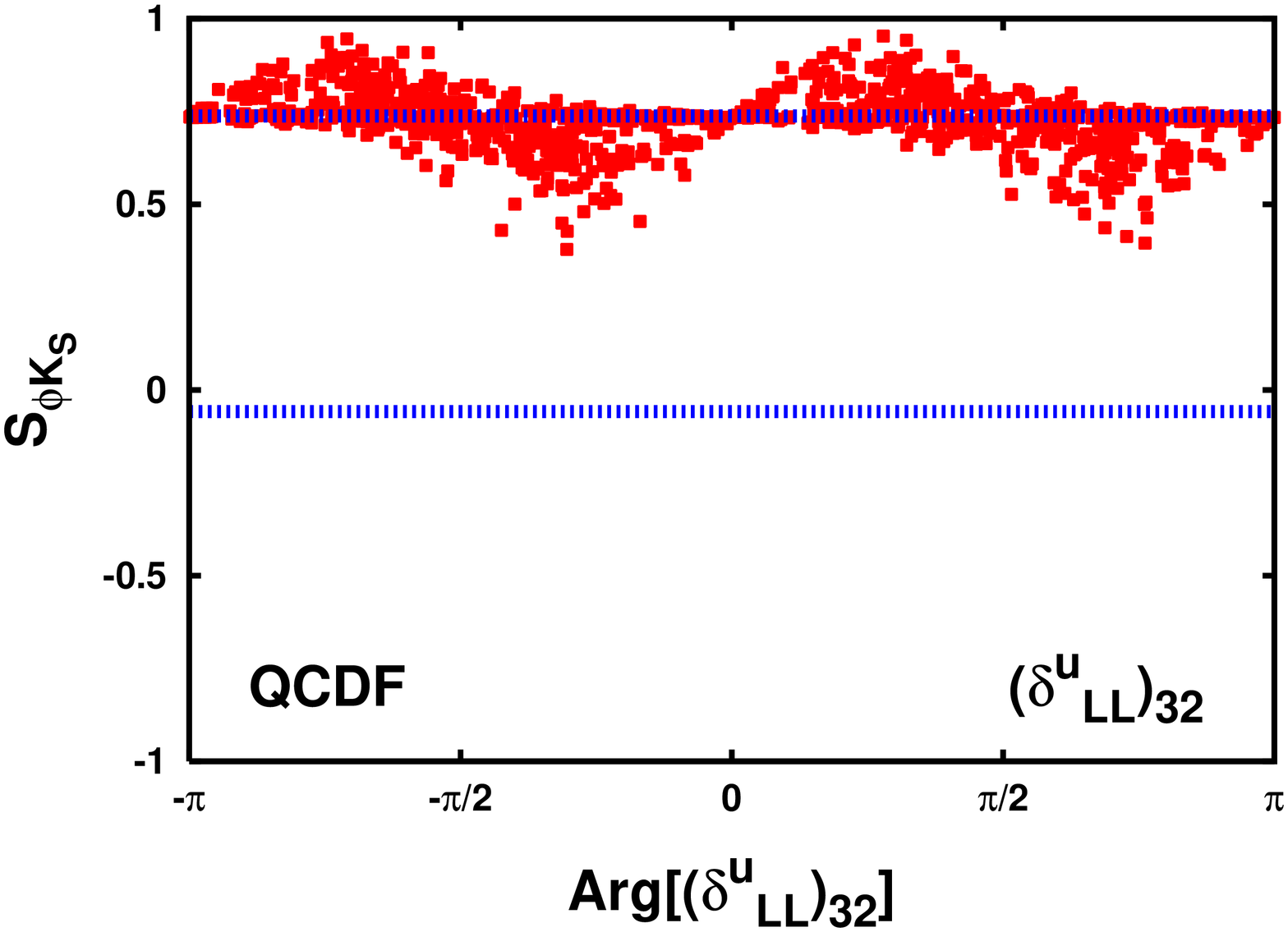, width=6truecm,
height=7truecm, angle=-90}}
\mbox{\epsfig{file=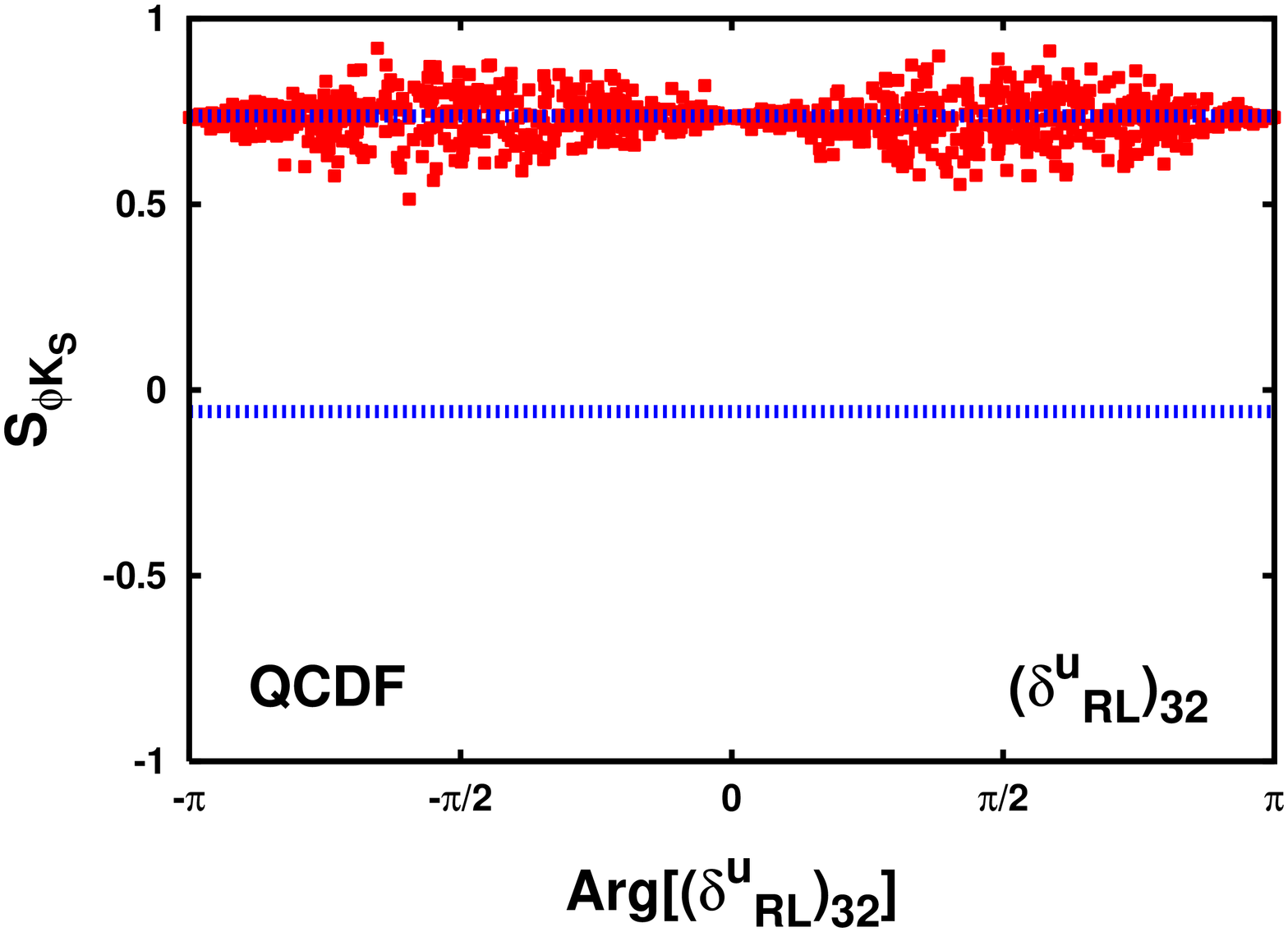, width=6truecm,
height=7truecm, angle=-90}}
\end{center}
\caption{\small As in Fig.~\ref{Fig41}, but for
$S_{\Phi K_S}$ as a function of arg[$\du{LL}{32}$]
(left) and  arg[$\du{RL}{32}$] (right) with chargino
contribution of one mass insertion .}
\label{Fig42}
\end{figure}

\begin{figure}[tpb]
\begin{center}
\mbox{\epsfig{file=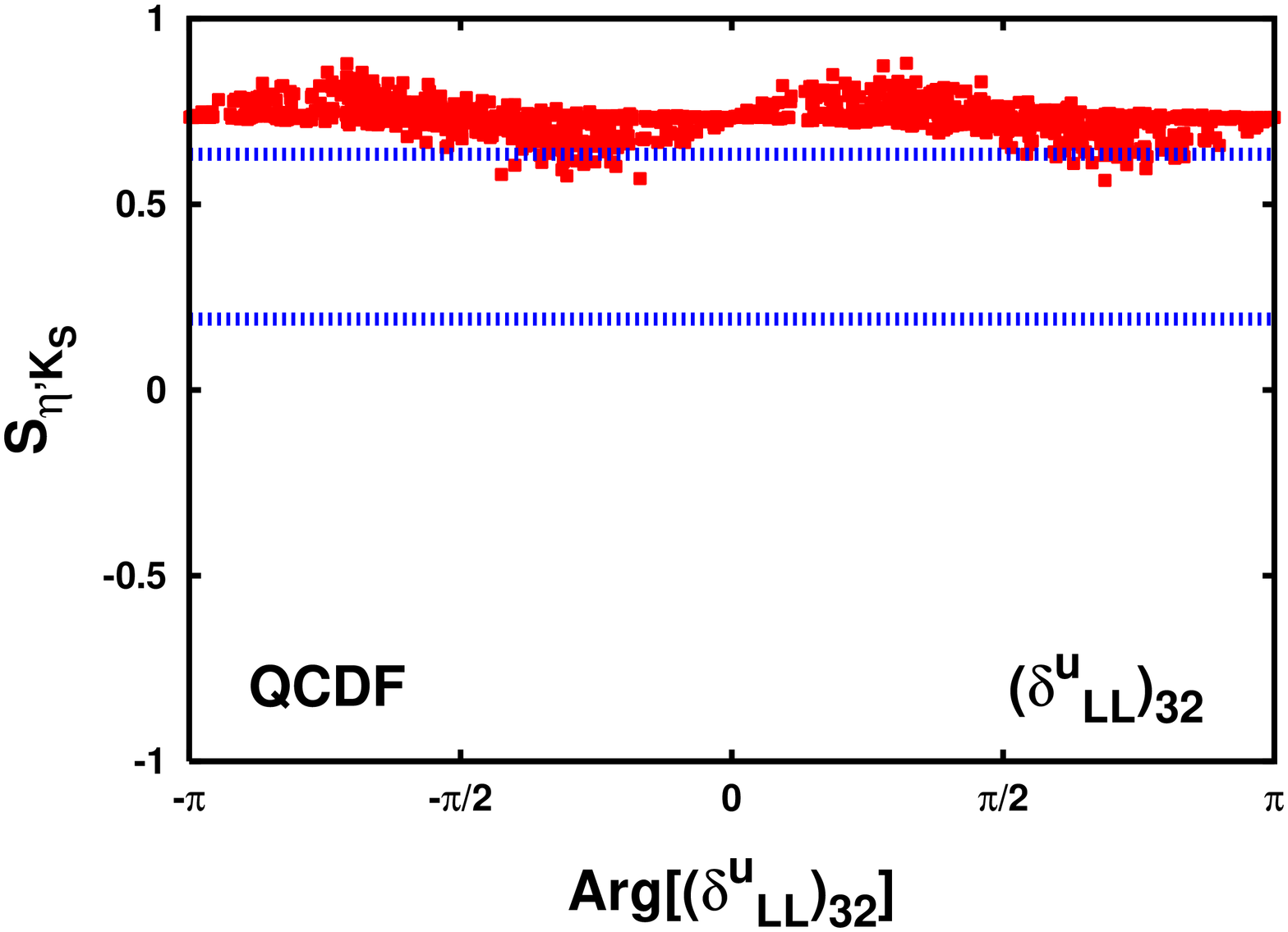, width=6truecm,
height=7truecm, angle=-90}}
\mbox{\epsfig{file=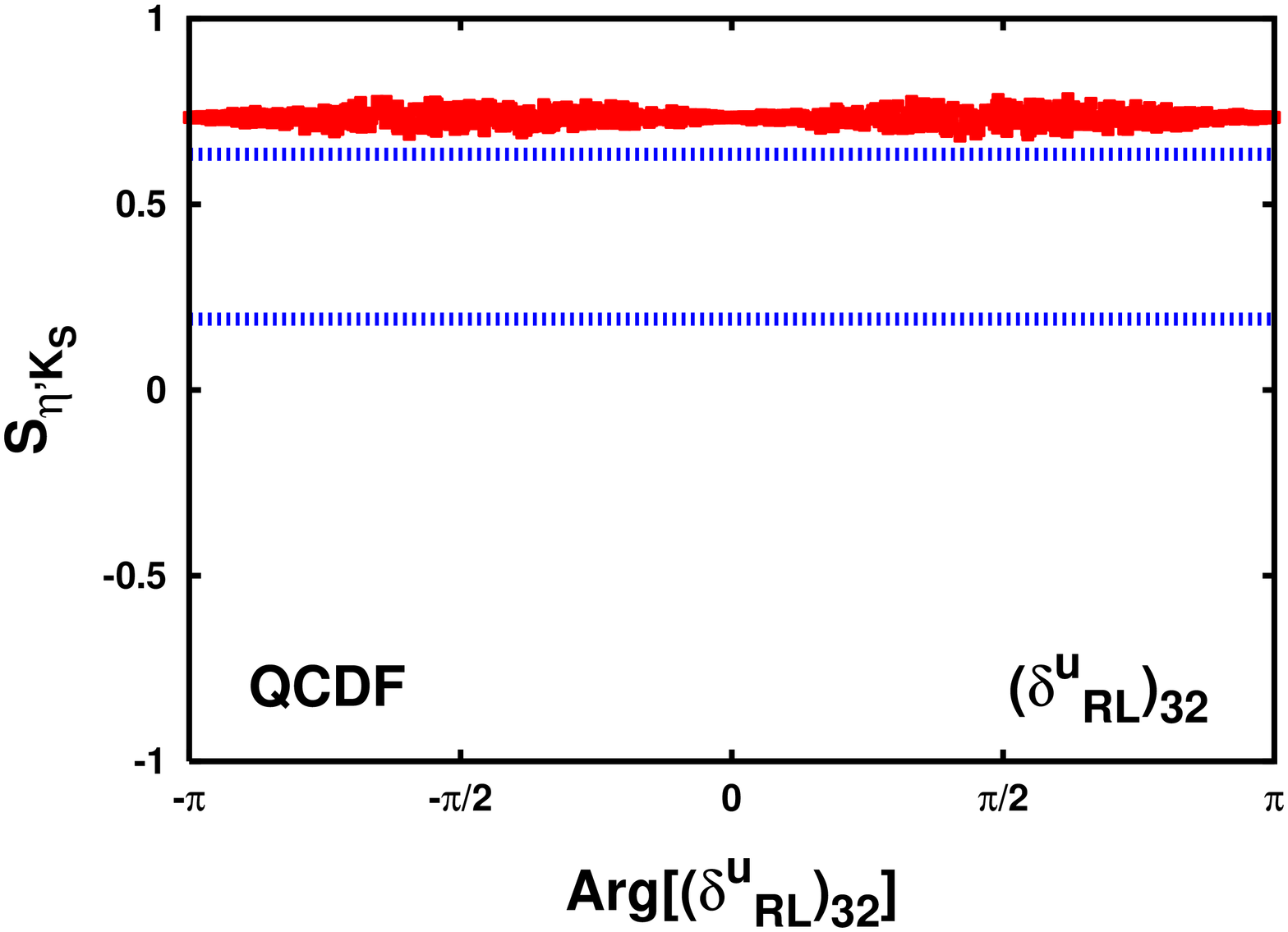, width=6truecm,
height=7truecm, angle=-90}}
\end{center}
\caption{\small As in Fig.~\ref{Fig42}, but for
$S_{\eta^{\prime} K_S}$.}
\label{Fig44}
\end{figure}

The chargino effects to $S_{\phi K_S}$ and $S_{\eta' K_S}$ 
\cite{CGHK,GHK} are
summarized in Fig.~\ref{Fig42} and Fig.~\ref{Fig44},
where $S_{\phi (\eta') K_S}$ is plotted versus the argument of the 
relevant chargino mass insertions, namely 
$\du{LL}{32}$ and $\du{RL}{32}$.

As in the gluino dominated scenario, we have scanned over the 
relevant SUSY parameter space, 
in particular, the average squark mass $\tilde{m}$, 
weak gaugino mass $M_2$, the $\mu$ term, 
and the light right stop mass $\tilde{m}_{\tilde{t}_R}$. 
Also $\tan{\beta}=40$ has been assumed and we take into 
account the present experimental bounds on SUSY spectra, in particular
$\tilde{m} > 300$ GeV, the lightest chargino mass $M_{\chi}>90$ GeV, 
and $\tilde{m}_{\tilde{t}_R} \ge 150$ GeV. As in the gluino case,
we scan over the real and imaginary part of 
the mass insertions $\du{LL}{32}$ and 
$\du{RL}{32}$, by considering the constraints on 
BR(\bsg) and \bbar mixing at 95\% C.L.. 
The \bsg constraints 
impose stringent bounds on $\du{LL}{32}$, specially at large $\tan{\beta}$
\cite{CGHK}.
Finally, as in the other plots, we scanned over the QCDF free parameters
$\rho_{A,H}<2 $ and $0<\phi_{A,H}<2\pi$.

The reason why extensive regions of negative values of $S_{\phi K_S}$
are excluded here, is only due to the \bsg constraints \cite{CGHK}.
Indeed, 
the inclusion of $\du{LL}{32}$ mass insertion 
can generate large and negative values of $S_{\phi K_S}$,
by means of chargino contributions to chromo-magnetic operator $Q_{8g}$
which are enhanced by terms of order $m_{\chi^{\pm}}/m_b$.
However, contrary to the gluino scenario, the ratio $|C_{8g}/C_{7\gamma}|$ 
is not enhanced by color factors and large contributions to
$C_{8g}$ leave unavoidably to the breaking of \bsg constraints.


\vspace{0.1cm}

\begin{figure}[tpb]
\begin{center}
\dofourfigs{3.1in}{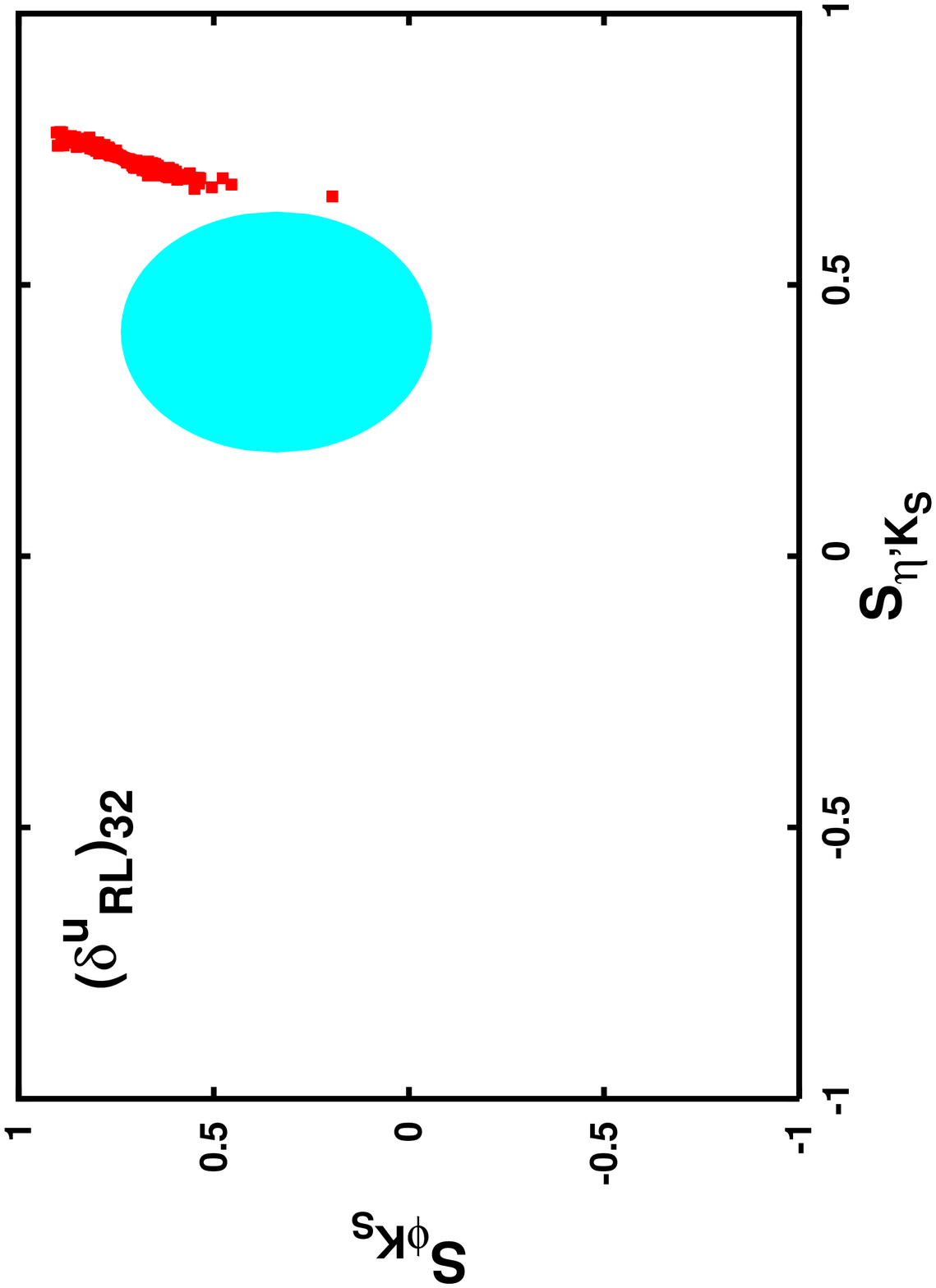}{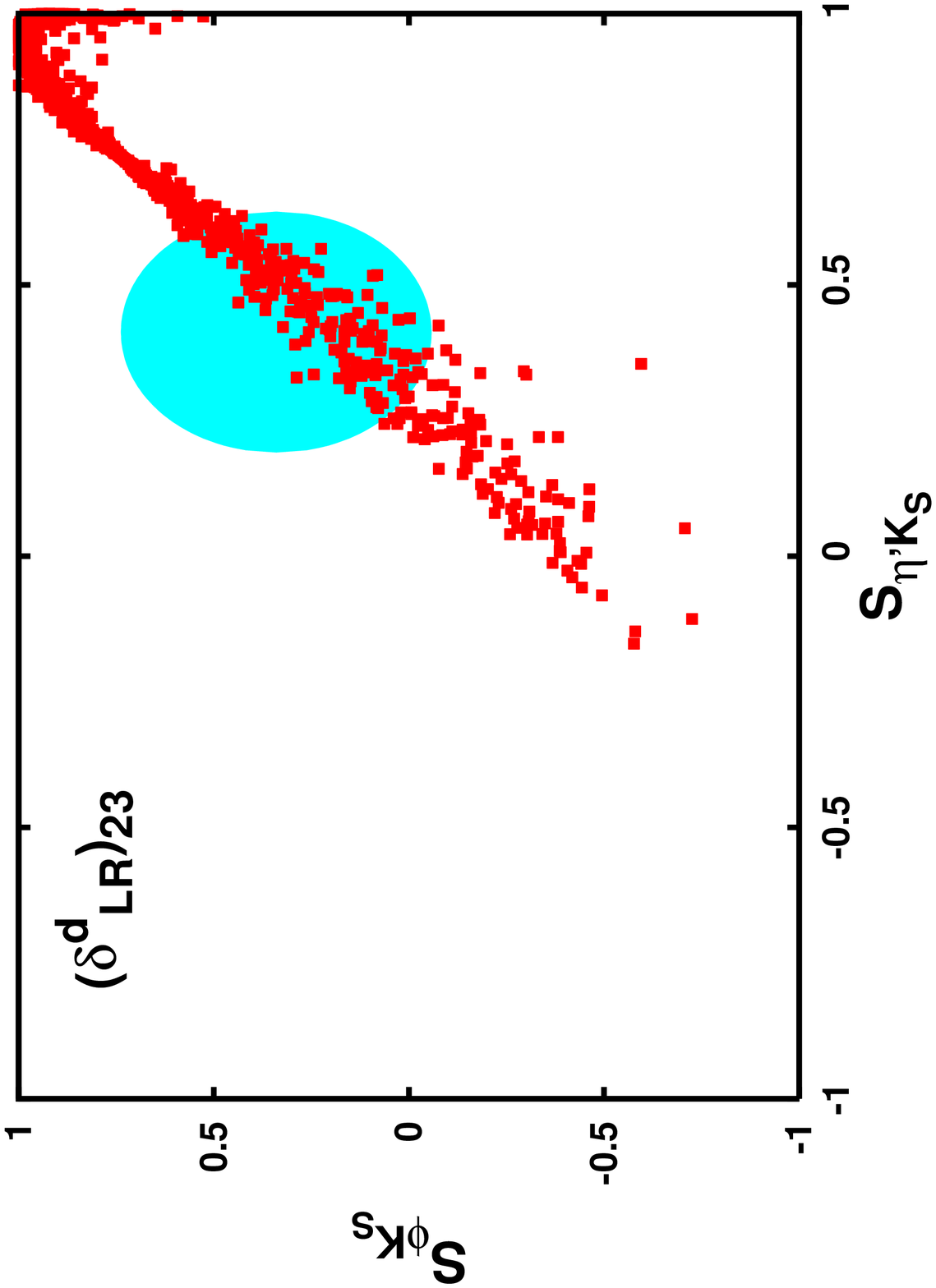}
{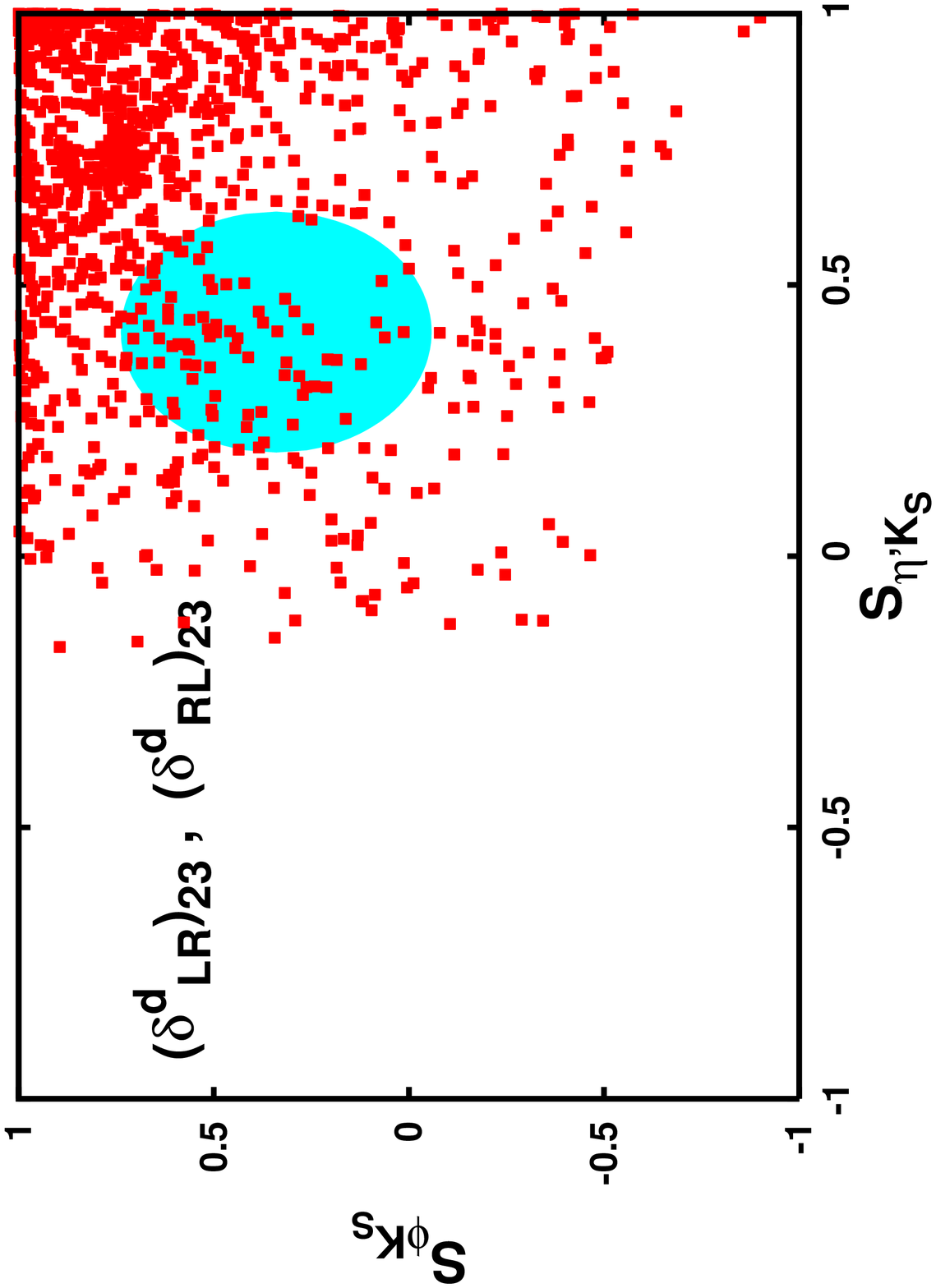}{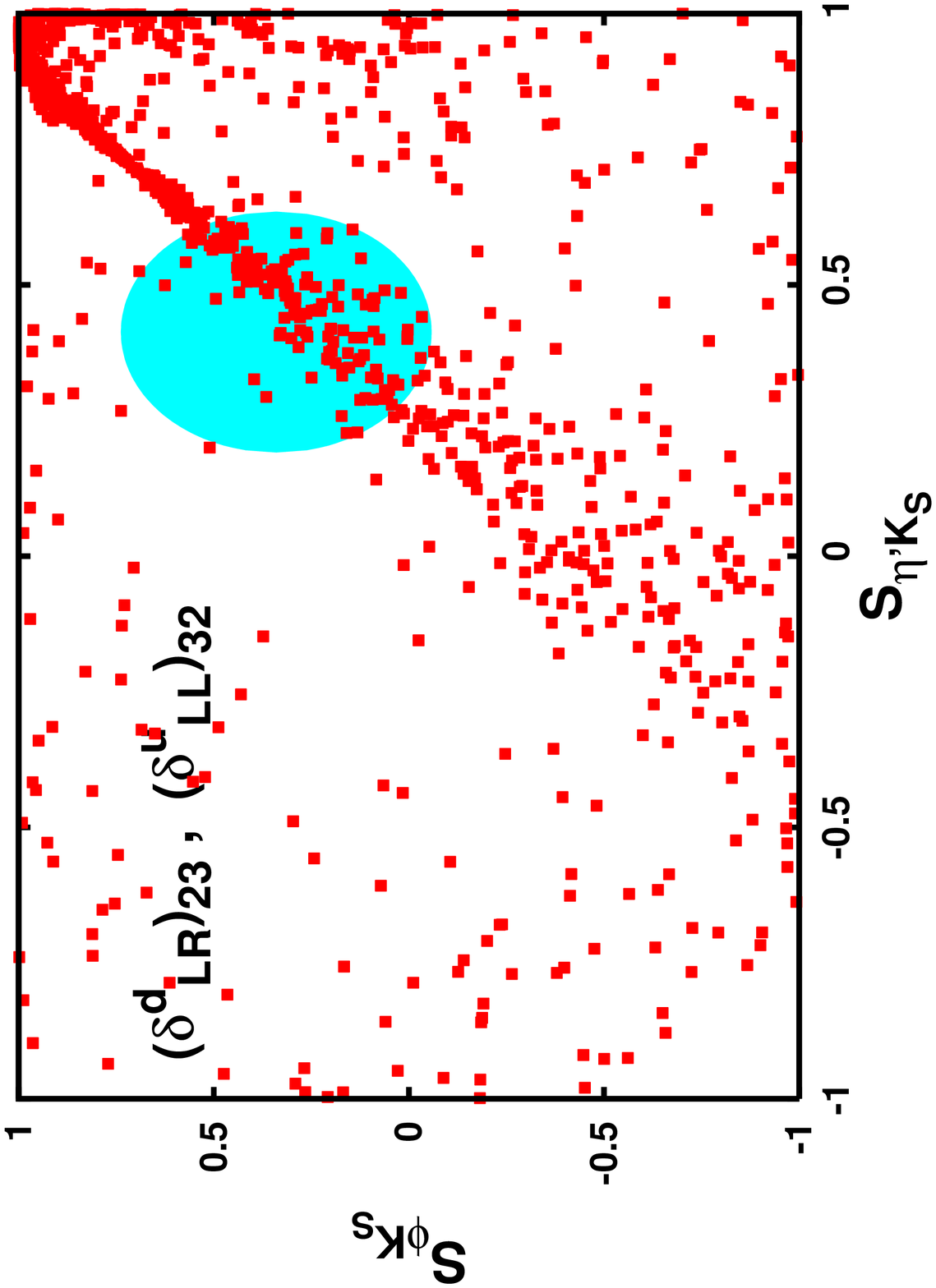}
\end{center}
\caption{\small Above: Correlation of 
asymmetries $S_{\Phi  K_S}$ versus 
$S_{\eta^{\prime} K_S}$ with the contribution of
one mass insertion $\du{RL}{32}$ (left) and 
$\dd{LR}{23}$ (right), for chargino (left) 
and gluino (right) exchanges. 
Region inside the ellipse corresponds to the allowed experimental ranges
at $2\sigma$ level. Below: as previously, but for 
arg[$\dd{LR}{23}$]=arg[$\dd{RL}{23}$] (left)
and 
arg[$\dd{LR}{23}$]=arg[$\du{LL}{32}$] (right),
with the contribution of two mass insertions
$\dd{LR}{23}$ \& $\dd{RL}{23}$ (left) for gluino exchanges,
and $\dd{LR}{23}$ \& $\du{LL}{32}$ (right)
for both gluino and chargino exchanges.}
\label{Fig45}
\end{figure}

\begin{figure}[tpb]
\begin{center}
\dofigs{3.1in}{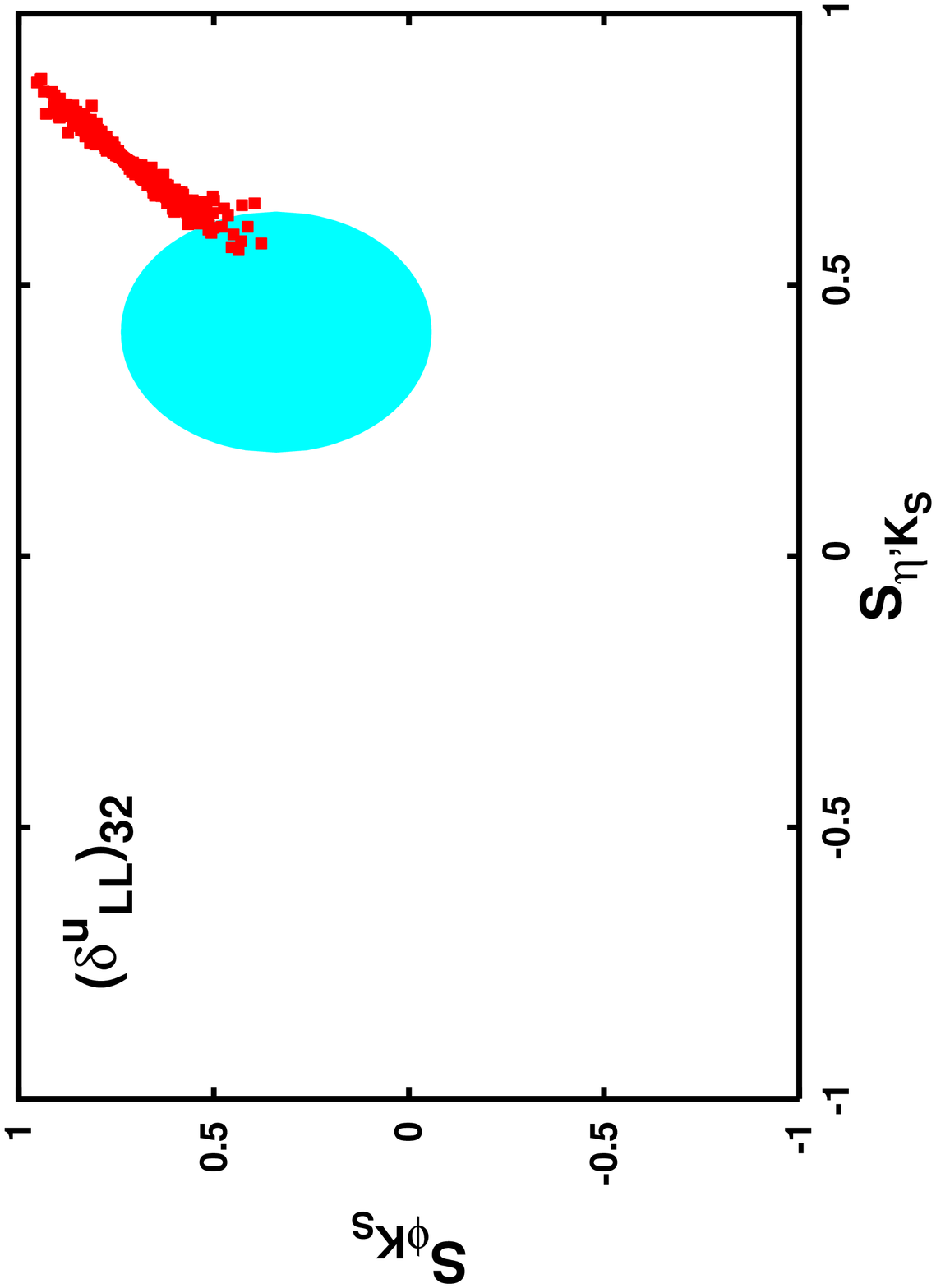}{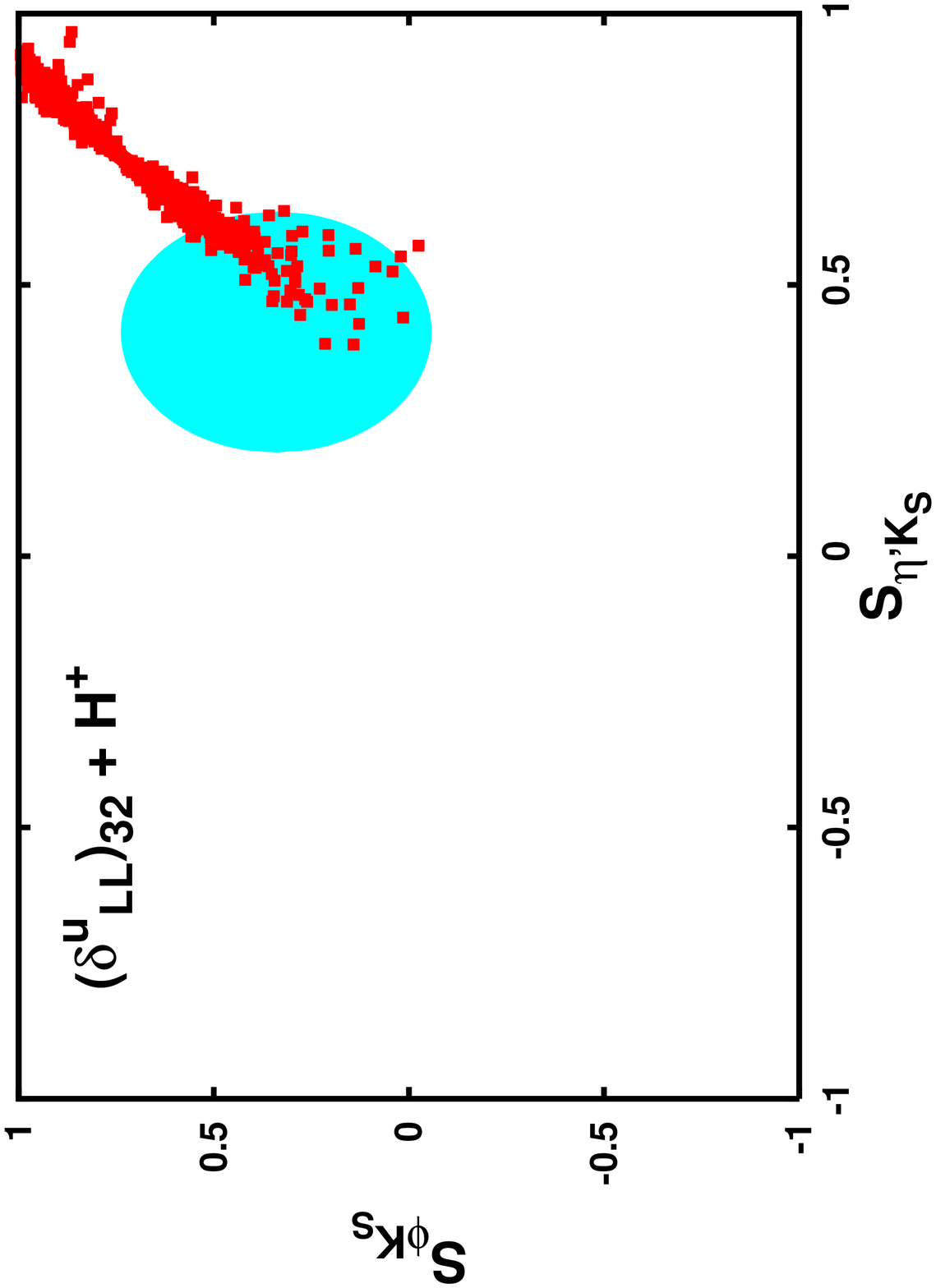}
\end{center}
\caption{\small 
Correlation of 
asymmetries $S_{\Phi  K_S}$ versus 
$S_{\eta^{\prime} K_S}$ for chargino contribution with 
single mass insertion $\du{LL}{32}$.
In the right plot the effect of a charged Higgs exchange, with mass
$m_H=200$ GeV and $\tan{\beta}=40$, has been taken into account.}
\label{FigH3}
\end{figure}

We plot
in Figs.~\ref{Fig45} the correlations between 
$S_{\phi K_S}$ versus $S_{\eta^{\prime} K_S}$ for both 
chargino and 
gluino in QCDF. For illustrative purposes, 
in all figures analyzing correlations,  we colored the area of 
the ellipse corresponding to the allowed experimental range
at $2\sigma$ level.\footnote{ All ellipses here
have axes of length $4\sigma$. As a first approximation, no correlation
between the expectation values of the two observables have been assumed.}

In Fig.~\ref{FigH3} 
the impact of a light charged Higgs in chargino 
exchanges is presented, when a charged Higgs with mass $m_H=200$ GeV 
and $\tan{\beta}=40$ has been taken into account.
The effects of charged Higgs exchange in 
the case of $\du{RL}{32}$ mass insertion are negligible,
as we expect from the fact that terms proportional to 
$\du{RL}{32}$ in $b\to s \gamma$ and $b\to s g$ amplitudes are not enhanced by 
$\tan{\beta}$.
On the other hand, in gluino exchanges with $\dd{LR}{23}$ or
$\dd{LL}{23}$, the most conspicuous 
effect of charged Higgs contribution is in populating the area outside 
the allowed experimental region. This
is due to a destructive interference with $b\to s\gamma$ amplitude, 
which relaxes the $b\to s\gamma$ constraints.
The most relevant effect of a charged Higgs exchange is in the 
scenario of chargino exchanges with $\du{LL}{32}$ mass insertion. In this case,
as can be seen from Fig.~\ref{FigH3}, a strong destructive interference with 
$b\to s\gamma$ amplitude can relax the $b\to s\gamma$ constraints in the 
right direction, allowing chargino predictions 
to fit inside the experimental region.
Moreover, we have checked that, for $\tan{\beta}=40$, 
charged Higgs heavier than approximately 600 GeV cannot affect the CP
asymmetries significantly.

%
\section{Direct CP asymmetry in $b \to s \gamma$ versus 
$S_{\phi(\eta^{\prime}) K_S}$}
%

Next we present the correlation for SUSY predictions
between the direct CP asymmetry $A_{CP}(b\to s \gamma)$ in \bsg decay
and the other ones in $B\to \phi (\eta^{\prime}) K_S$.
The CP asymmetry in \bsg is measured 
in the inclusive radiative decay of $B \to X_s \gamma$ by the quantity 
\bea
A_{CP}(b\to s \gamma) = \frac{\Gamma(\bar{B} \to X_s \gamma) - \Gamma(B \to
X_{\bar{s}}
\gamma)}{\Gamma(\bar{B} \to X_s \gamma) + \Gamma(B \to X_{\bar{s}}
\gamma)}.
\label{Absg}
\eea
The SM prediction for $A_{CP}(b\to s \gamma)$ is very small, less than
$1\%$ in magnitude, but known with high precision \cite{ACPbsg}.
Indeed, inclusive decay rates of B mesons are free from large
theoretical uncertainties since they can be reliably calculated in QCD using 
the OPE.
Thus, the observation of sizeable effects in 
$A_{CP}(b\to s \gamma)$ would be a clean signal of new physics.
In particular, large asymmetries are expected in models with enhanced 
chromo-magnetic dipole operators, like for instance supersymmetric models
\cite{ACPbsg}. 

\begin{figure}[tpb]
\begin{center}
\dofigs{3.1in}{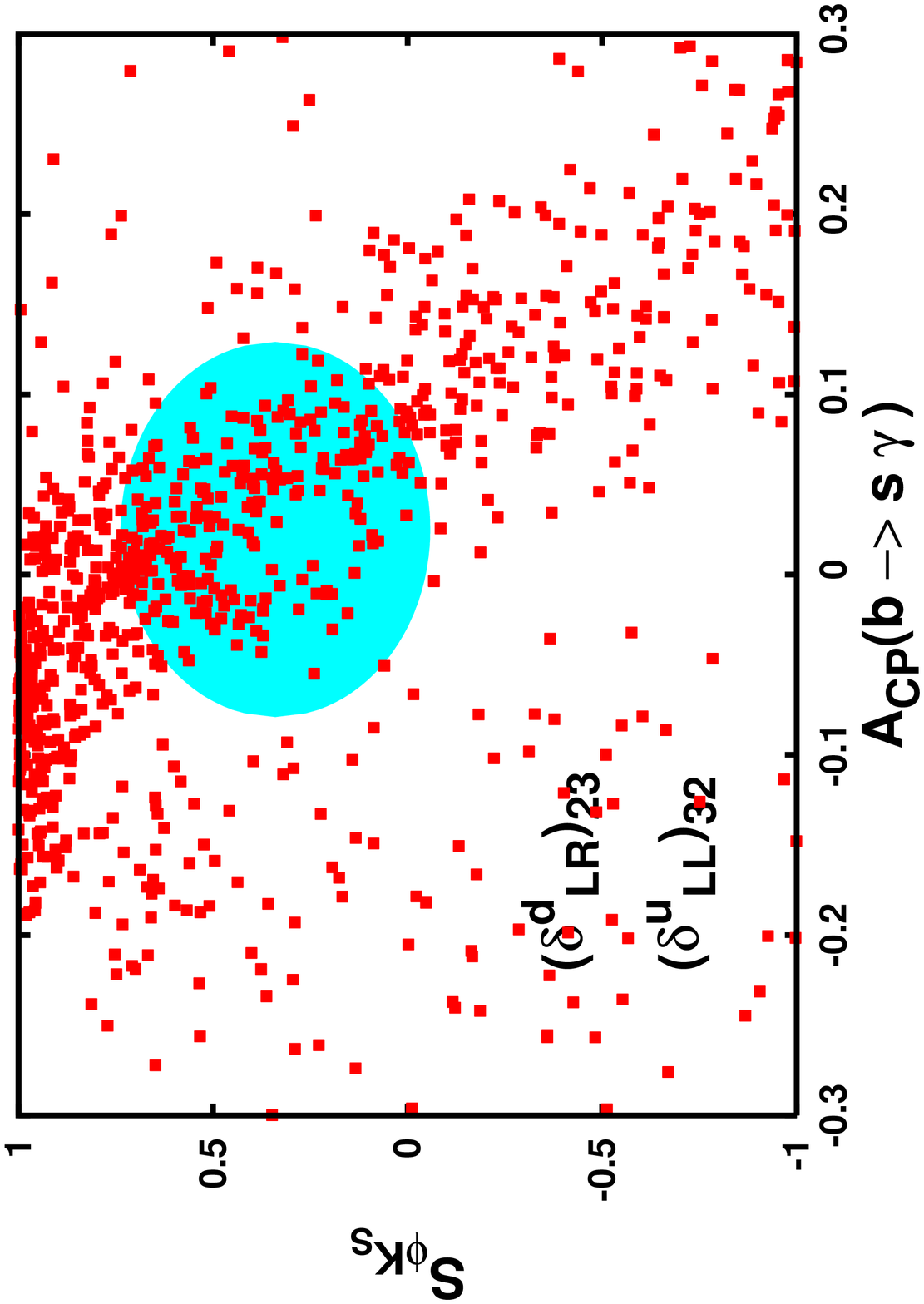}{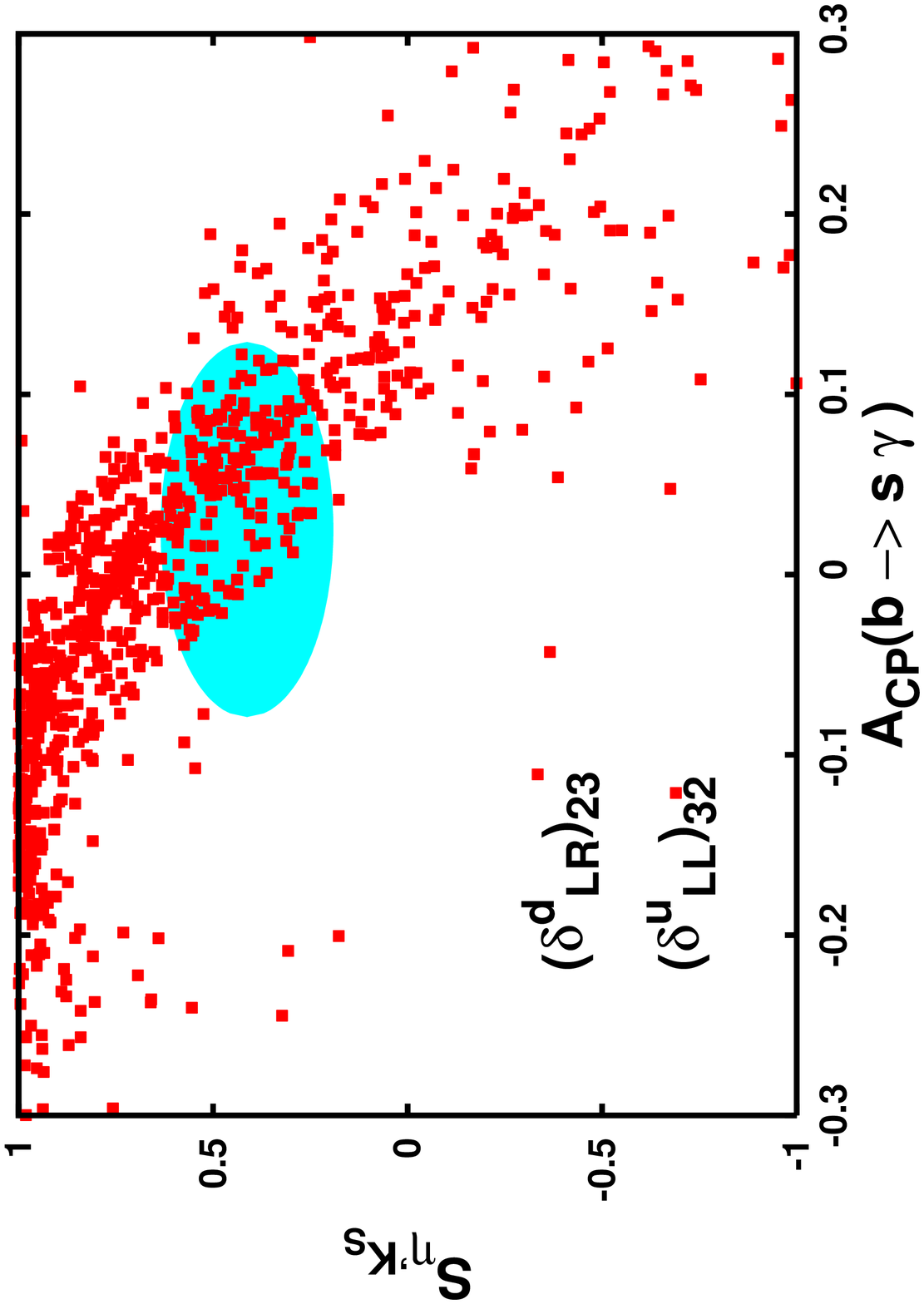}
\end{center}
\caption{\small Correlations of $S_{\phi K_S}$ versus $A_{CP}(b\to s
\gamma )$ (left) and
$S_{\eta^{\prime} K_S}$ versus $A_{CP}(b\to s \gamma )$
(right),
for gluino and chargino contributions with
mass insertions $\dd{LR}{23}$ and $\du{LL}{32}$ respectively.
}
\label{ABSG_DLR_ULL}
\end{figure}

In  Fig.~\ref{ABSG_DLR_ULL} we show our results for two
mass insertions $\dd{LR}{23}$ and $\du{LL}{32}$
with both gluino and chargino exchanges.
In this case we see that $S_{\phi K_S}$ constraints 
do not set any restriction on $A_{CP}(b\to s\gamma)$, and 
also large and positive values of $A_{CP}(b\to s\gamma)$ asymmetry 
can be achieved.
However, by imposing the constraints on
$S_{\eta^{\prime} K_S}$, see plot on the right side of Fig.~\ref{ABSG_DLR_ULL},
the region of negative $A_{CP}(b\to s\gamma)$ is disfavored in this scenario 
as well.

%
\section{Conclusions}
%
CP violation remains as one of the most interesting research topics
both theoretically and experimentally.  
Especially the $B$-meson decay modes seem to be ideally suited
for searching effects of new physics, since new interesting results
are mounting from the $B$-factories.  
For interpretation of those results, which get more accurate with
more statistics, also reducing theoretical uncertainties in the 
calculations is a big challenge.

The $B$-meson decays to $\phi K$, $\eta^{\prime} K$, and to
$X_S\gamma$ provide a clean window to the physics beyond the SM.
Here our analysis of the supersymmetric 
contributions to the CP asymmetries and the branching ratios of 
these processes in a model independent way has been reviewed.
The relevant SUSY contributions
in the $b\to s$ transitions, namely chargino and gluino exchanges 
in box and penguin diagrams, have been considered 
by using the mass insertion method.

Due to the stringent constraints from the experimental 
measurements of $BR(b\to s \gamma)$, the scenario with 
pure chargino exchanges cannot give large and negative 
values for CP asymmetry $S_{\phi K_S}$. 
It is, however, seen that charged Higgs may enhance the chargino
contributions substantially.
On the other hand, it is quite possible for gluino exchanges to account for
$S_{\phi K_S}$ and $S_{\eta^{\prime} K_S}$ at the same time.

We also discussed the correlations between the CP asymmetries of these 
processes and the direct CP asymmetry in $b\to s \gamma$ decay. 
In this case, we found that the general trend of 
SUSY models, satisfying all the experimental constraints, 
favors large and positive contributions 
to $b\to s \gamma$ asymmetry.

\vspace{0.5in}

\noindent
{\bf Acknowledgments} 

\vspace{0.2in} 
\noindent
KH thanks the German University in Cairo for hospitality and the
pleasant atmosphere of the 1st GUC Workshop on High Energy Physics.
The authors appreciate the financial support from the Academy of Finland
(project numbers 104368 and 54023).



\end{document}